%% file: oup-authoring-template_WG.tex
\documentclass[unnumsec,webpdf,contemporary,large]{oup-authoring-template}%

\graphicspath{{Fig/}}

\theoremstyle{thmstyleone}%
\theoremstyle{thmstyletwo}%
\theoremstyle{thmstylethree}%

\usepackage{amsmath,amsmath,amssymb}
\usepackage{mathtools}
\usepackage{subfiles}
\DeclareMathOperator*{\argmax}{arg\,max}
\begin{document}

\journaltitle{Briefings in Bioinformatics}
\DOI{DOI HERE}
\copyrightyear{2023}
\pubyear{2023}
\access{Advance Access Publication Date: Day Month Year}
\appnotes{Paper}

\firstpage{1}


\title[Artificial intelligence-aided protein engineering]{Artificial intelligence-aided protein engineering: from topological data analysis to deep protein language models}

\author[1]{Yuchi Qiu}
\author[1,2,3,$\ast$]{Guo-Wei Wei}


\address[1]{\orgdiv{Department of Mathematics}, \orgname{Michigan State University}, \orgaddress{\street{East Lansing}, \postcode{48824}, \state{MI}, \country{USA}}}
\address[2]{\orgdiv{Department of Biochemistry and Molecular Biology}, \orgname{Michigan State University}, \orgaddress{\street{East Lansing}, \postcode{48824}, \state{MI}, \country{USA}}}
\address[3]{\orgdiv{Department of Electrical and Computer Engineering}, \orgname{Michigan State University}, \orgaddress{\street{East Lansing}, \postcode{48824}, \state{MI}, \country{USA}}}

\corresp[$\ast$]{Corresponding author. \href{weig@msu.edu}{weig@msu.edu}}

\received{Date}{0}{Year}
\revised{Date}{0}{Year}
\accepted{Date}{0}{Year}



\abstract{Protein engineering is an emerging field in biotechnology that has the potential to revolutionize various areas, such as antibody design, drug discovery, food security, ecology, and more. However, the mutational space involved is too vast to be handled through experimental means alone. Leveraging accumulative protein databases, machine learning (ML) models, particularly those based on natural language processing (NLP), have considerably expedited protein engineering. Moreover, advances in topological data analysis (TDA) and artificial intelligence-based protein structure prediction, such as AlphaFold2, have made more powerful structure-based ML-assisted protein engineering strategies possible. This review aims to offer a comprehensive, systematic, and indispensable set of methodological components, including TDA and NLP, for protein engineering and to facilitate their future development.}


\maketitle
\section{Key points}
\begin{itemize}
\item Machine learning and deep learning techniques are revolutionizing protein engineering.
\item Topological data analysis enables advanced structure-based machine learning-assisted protein engineering approaches.
\item Deep protein language models extract critical evolutionary information from large-scale sequence databases.
\end{itemize}
\section{Introduction}
Protein engineering aims to design and discover proteins with desirable functions, such as improving the phenotype of living organisms, enhancing enzyme catalysis, and boosting antibody efficacy \cite{narayanan2021machine}. It has tremendous impacts on drug discovery, enzyme development and applications, the development of biosensors, diagnostics, and other biotechnology, as well as understanding the fundamental principles of the protein structure-function relationship and achieving environmental sustainability and diversity. Protein engineering has the potential to continue to drive innovation and improve our lives in the future.

Two traditional protein engineering approaches include directed evolution \cite{arnold1998design} and rational design \cite{karplus2005molecular, boyken2016novo}. Directed evolution is a process used to create proteins or enzymes with improved or novel functions \cite{romero2009exploring}. The method involves introducing mutations into the genetic code of a target protein and screening the resulting variants for improved function. The process is "directed" because it is guided by the desired outcome, such as increased activity, stability, specificity, binding affinity, and fitness. Rational design involves using knowledge of protein structure and function to engineer desirable specific changes to the protein sequence and/or structure \cite{boyken2016novo, bhardwaj2016accurate}. Both approaches resort to experimental screening of astronomically large mutational space, i.e., $20^N$ for protein of $N$ amino acid residues, which is expensive, time-consuming, and intractable \cite{pierce2002protein}. As a result, only a small fraction of the mutational space can be explored experimentally even with the most advanced high-throughput screening technology.

Recently, data-driven machine learning has emerged as a new approach for directed evolution and protein engineering \cite{siedhoff2020machine, mazurenko2019machine}. Machine learning-assisted protein engineering (MLPE) refers to the use of machine learning models and techniques to improve the efficiency and effectiveness of protein engineering. MLPE not only reduces the cost and expedites the process of protein engineering, but also optimizes the screening and selection of protein variants \cite{diaz2023using}, leading to the higher efficiency and productivity. Specifically, by using machine learning to analyze and predict the effects of mutations on protein function, researchers can rapidly generate and test large numbers of variants, which establish the protein-to-fitness map (i.e., fitness landscape) from sparsely sampled experimental data \cite{wittmann2021advances, yang2019machine}. This approach accelerates the process of protein engineering.

The process of data-driven MLPE typically involves several elements, including data collection and preprocessing, model design, feature extraction and selection, algorithm selection and design, model training and validation, experimental validation, and iterative model optimization. Driven by technological advancements in high-throughput sequencing and screening technologies, there has been a substantial accumulation of general-purpose experimental datasets on protein sequences, structures, and functions \cite{berman2000protein, uniprot2021uniprot}. These datasets, along with numerous protein-engineering specific deep mutational scanning (DMS) libraries \cite{notin2022tranception}, provide valuable resources for machine learning training and validation.    

Data representation and feature extraction are crucial steps in the design of machine learning models, as they help to reduce the complexity of biological data and enable more effective model training and prediction. There are several typical types of feature embedding methods, including sequence-based, structure-based \cite{cang2018integration, wang2020topology}, physics-based \cite{schymkowitz2005foldx, leman2020macromolecular}, and hybrid methods \cite{qiu2023persistent}. Among them, sequence-based embeddings have been dominant due to the success of various natural language processing (NLP) methods such as long short-term memory (LSTM) \cite{alley2019unified}, autoencoders \cite{riesselman2018deep}, and Transformers \cite{rives2021biological}, which allow unsupervised pre-training on large-scale sequence data. Structure-based embeddings take advantage of existing protein three-dimensional (3D) structures in the Protein Data Bank (PDB) \cite{berman2000protein} and advanced structure predictions such as AlphaFold2 \cite{jumper2021highly}. These methods further exploit advanced mathematical tools, such as topological data analysis (TDA) \cite{edelsbrunner2010computational, zomorodian2005computing}, differential geometry \cite{nguyen2019dg,wee2021ollivier}, or graph approaches \cite{nguyen2019agl}. Physics-based methods utilize physical models, such as density functional theory \cite{ryczko2019deep}, molecular mechanics \cite{butler2018machine}, Poisson-Boltzmann model \cite{chen2021mlimc}, etc. While these methods are highly interpretable, their performance often depends on model parametrization. Hybrid methods may select a combination of two or more types of features.

The designs and selections of MLPE algorithms depend on the availability of data and efficiency of experiments. In real-world scenarios, where smaller training datasets are prevalent, simpler machine learning algorithms such as support vector machines and ensemble methods are often employed for small training datasets, which is often the case in real scenarios. In contrast, deep neural networks  are more suitable for larger training datasets. Although regression tasks are typically used to distinguish one set of mutations from another \cite{siedhoff2020machine}, unsupervised zero-shot learning methods can also be utilized to address scenarios with limited data availability \cite{hsu2022learning, wittmann2021informed}. The iterative interplay between experiments and models is another crucial component in MLPE by iteratively screening new data to refine the models. Consequently, the selection of an appropriate MLPE model is influenced by factors like experimental frequency and throughput. This iterative refinement process enables MLPE to deliver optimized protein engineering outcomes.

\begin{figure}[!t]%
	\centering
	\includegraphics[width=3in]{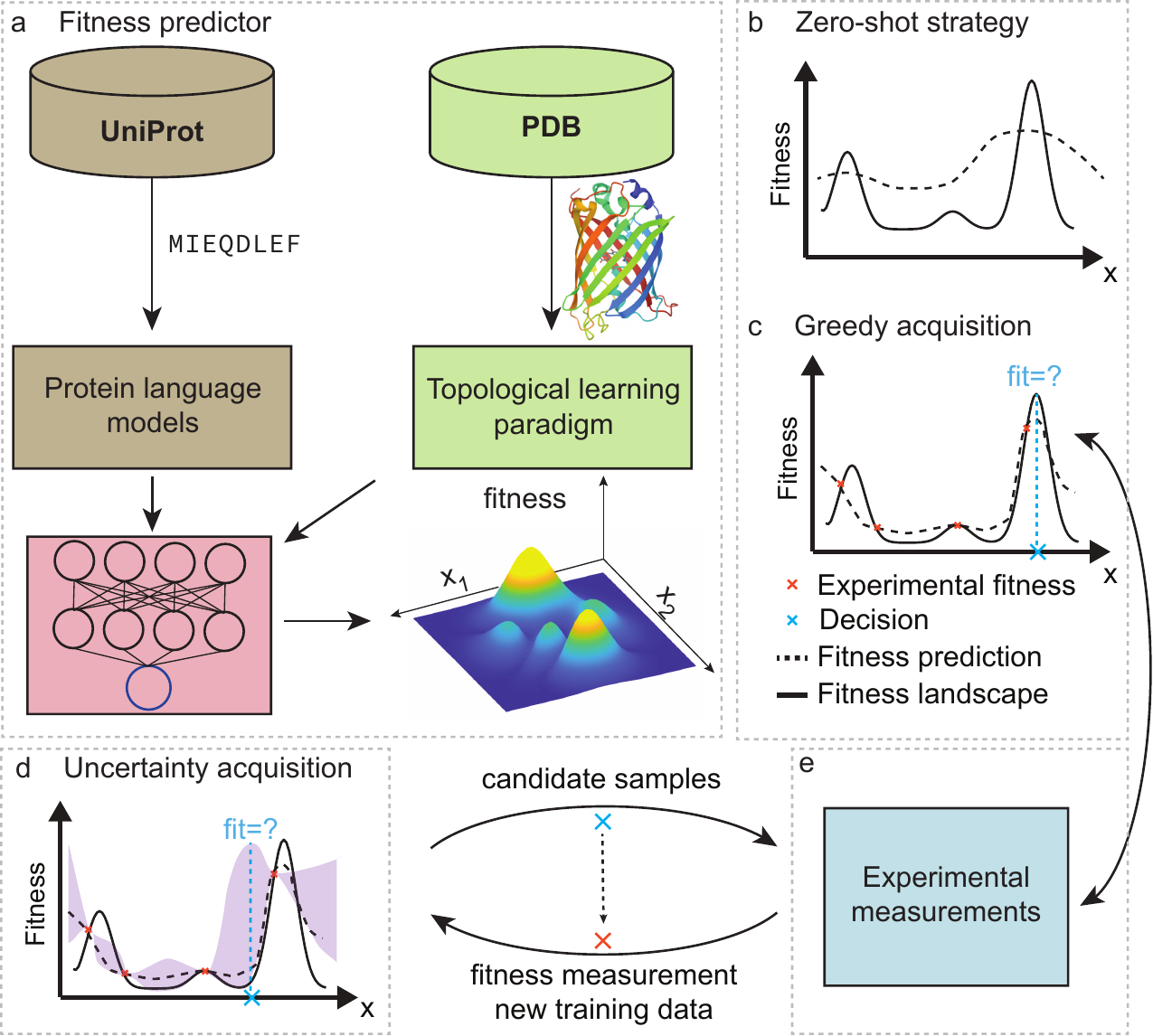}
	\captionsetup{width=14cm}
	\caption{\textbf{Machine learning-assisted protein engineering (MLPE).} (a). Machine learning models build fitness predictor using structure and sequence protein data. (b). Zero-shot predictors navigate fitness landscape without labeled data. (c). Greedy acquisition exploits fitness using fitness predictions. (d). Uncertainty acquisition balances exploitation and exploration. The example shows a Gaussian upper confidence bound (UCB) acquisition. (e). Experimental measurements query fitness of candidate proteins in sequential optimization.  }
\label{fig2}
\end{figure}

MLPE has the potential to significantly accelerate the development of new and improved proteins, revolutionizing numerous areas of science and technology (\autoref{fig2}). 
Despite considerable advances in MLPE, challenges remain in many aspects, such as data preprocessing, feature extraction, integration with advanced algorithms, and iterative optimization through experimental validation. This review examines published works and offers insights into these technical advances. We place particular emphasis on the advanced mathematical TDA approaches, aiming to make them accessible to general readers. Furthermore, we review current advanced NLP-based models and efficient MLPE approaches. Last, we discuss potential future directions in the field.

\section{Sequence-based deep protein language models}
\label{sec_PLM}

In artificial intelligence, natural language processing (NLP) has recently gained much attention for representing and analyzing human language computationally \cite{khurana2023natural}. NLP covers a wide range of tasks, including language translation, sentiment analysis, chatbot development, speech recognition, and information extraction,  among others. The development and advancement of various machine learning models have been instrumental in tackling the complex challenges posed by NLP tasks.

Similar to human language, the primary structure of a protein is also represented by a string of amino acids, with 20 canonical amino acids. The analogy between protein sequences and human languages has inspired the development of computational methods for analyzing and understanding proteins using models adopted from NLP (\autoref{fig2}a). The self-supervised sequence-based protein language models have been applied to study the underlying patterns and relationships within protein sequences, predict their structural and functional properties, and facilitate protein engineering. These language models are pretrained on a given data allowing to model protein properties for each given protein. There are two major types of protein language models utilizing different resources of protein data \cite{hsu2022learning} (\autoref{table_seq}). The first one is the local evolutionary models which focus on homologs of the target protein such as multiple sequence alignments (MSAs) to learn the evolutionary information from the mostly related mutations. The second one is the global evolutionary models which learn from large protein sequence databases such as UniProt \cite{uniprot2021uniprot} and Pfam \cite{el2019pfam}. 

{\small 
\begin{table*}[h]
\centering
\begin{tabular}{|p{2.6cm}|p{2.5cm}|p{1.2cm}|p{0.7cm}|p{1cm}|p{2.2cm}|p{2cm}|p{1.4cm}|}
\hline
\multirow{2}{*}{\textbf{Model} }        & \multirow{2}{*}{\textbf{Architecture}}& \multirow{2}{*}{\textbf{Max len}}& \multirow{2}{*}{\textbf{Dim}}& \multirow{2}{*}{\textbf{\# para}} &\multicolumn{2}{|c|}{\textbf{Pretrained data}}& \multirow{2}{*}{Time$^1$}\\ 
\cline{6-7}
& &&&& \textbf{Source} &\textbf{Size}&      \\ \hline
\multicolumn{8}{|c|}{\textbf{Local Models}} \\ \hline
Profile HMMs \cite{shihab2013predicting}                   & Hidden Markov&-- &--          &--&MSAs &--&Oct 2012\\ \hline
EvMutation \cite{hopf2017mutation}           & Potts Models  &--&--&--&MSAs  &-- &Jan 2017               \\ \hline
MSA Transformer \cite{rao2021msa}        & Transformer&1024&768&100M     & UniRef50 \cite{uniprot2021uniprot} &26M    &Feb 2021           \\ \hline
DeepSequence \cite{riesselman2018deep}           & VAEs&--&--&--&MSAs&--&Dec 2017 \\ \hline
EVE \cite{frazer2021disease}                    & Bayesian VAEs &--&--&--&MSAs &--&Oct 2021 \\ \hline
\multicolumn{8}{|c|}{\textbf{Global Models}} \\ \hline
TAPE ResNet \cite{rao2019evaluating}                 & ResNet&1024 &256&38M  & Pfam \cite{el2019pfam} &31M& Jun 2019             \\ \hline
TAPE LSTM \cite{rao2019evaluating} 	          & LSTM&1024 &2048&38M& Pfam \cite{el2019pfam} &31M& Jun 2019\\ \hline
TAPE Transformer \cite{rao2019evaluating}     & Transformer&1024 &512&38M   & Pfam \cite{el2019pfam} &31M& Jun 2019            \\ \hline
Bepler \cite{bepler2018learning} 		              & LSTM &512  &100&22M& Pfam \cite{el2019pfam} &31M   &Feb 2019                       \\ \hline
UniRep \cite{alley2019unified}                 & LSTM &512 &1900&18M&UniRef50 \cite{uniprot2021uniprot}&      24M   &Mar 2019                \\ \hline
eUniRep \cite{biswas2021low} & LSTM   &512&1900&18M&UniRef50 \cite{uniprot2021uniprot}; MSAs &      24M                     &Jan 2020                      \\ \hline
ESM-1b \cite{rives2021biological}                    & Transformer &1024&1280&650M&UniRef50 \cite{uniprot2021uniprot} &250M& Dec 2020                   \\ \hline
ESM-1v \cite{meier2021language}                    & Transformer  &1024&1280&650M&                  UniRef90 \cite{uniprot2021uniprot} &98M&Jul 2021 \\ \hline
ESM-IF1 \cite{hsu2022esm}&Transformer&--&512&124M&UniRef50 \cite{uniprot2021uniprot}; CATH \cite{orengo1997cath}&12M sequences; 16K structures&Sep 2022\\ \hline
ProGen \cite{madani2023large}                 & Transformer&512 &--&1.2B&UniParc \cite{uniprot2021uniprot}; UniprotKB \cite{uniprot2021uniprot}; Pfam \cite{el2019pfam}; NCBI Taxonomy \cite{federhen2012ncbi} &281M&               Jul 2021 \\ \hline
ProteinBERT \cite{brandes2022proteinbert}            & Transformer &1024&--&16M&UniRef90 \cite{uniprot2021uniprot}&106M               &May 2021     \\ \hline
Tranception \cite{notin2022tranception}            & Transformer     &1024&1280&700M              &UniRef100 \cite{uniprot2021uniprot}&250M&May 2022   \\ \hline
ESM-2 \cite{lin2023evolutionary}           & Transformer &1024  &5120&15B&    UniRef90                 \cite{uniprot2021uniprot}&65M&Oct 2022   \\ \hline
\end{tabular}
\caption{Summary of protein language models. \# para: number of parameters which are only provided for deep learning models. Max len: maximum length of input sequence. Dim: latent space dimension. Size: pre-trained data size where it refers to number of sequences without specification except MSA transformer includes 26 millions of MSAs. K: thousands; M: millions; B: billions. $^1$: Time for the first preprint. The input data size, hidden layer dimension, and number of parameters are only provided for global models. }
\label{table_seq}
\end{table*}
}
\subsection{Local evolutionary models}

To train a local evolutionary model, MSAs search strategies such as jackhmmer \cite{eddy2011accelerated} and EvCouplings \cite{hopf2019evcouplings} are first employed. Taking  MSAs as inputs, local evolutionary models learn the probabilistic distribution of mutations for a target protein. Probabilistic models, including Hidden Markov Models (HMMs) \cite{rabiner1989tutorial, shihab2013predicting} and Potts-based models \cite{hopf2017mutation},   are popular in modeling mutational effects.  Transformer models have been introduced to learn distribution from MSAs. The MSA Transformer \cite{rao2021msa} introduces a row- and column-attention mechanism. Recent years, variational autoencoders (VAEs) \cite{kingma2013auto} serve as the alternate to model MSAs by including the dependency between residues and aligning all sequences to a probability distribution. The VAE model DeepSequence \cite{riesselman2018deep} and the Bayesian VAE model EVE \cite{frazer2021disease} exhibit excellent performance in modeling mutational effects \cite{livesey2020using, hsu2022learning, qiu2023persistent}.  

\subsection{Global evolutionary models}

With  large-size  data, global evolutionary models usually adopt the large NLP models. Convolutional Neural networks (CNNs) \cite{kim-2014-convolutional} models and residual network (ResNet) \cite{he2016deep} have been employed for protein sequence analysis \cite{rao2019evaluating}. Large-scale models, such as long short-term memory (LSTM)  \cite{hochreiter1997long}, have also gained popularity as seen in Bepler \cite{bepler2018learning}, UniRep \cite{alley2019unified}, and eUniRep \cite{biswas2021low}. In recent years, the Transformer architecture has achieved state-of-the-art performance in NLP by introducing the attention mechanism and the self-supervised learning via the masked filling training strategy \cite{vaswani2017attention, devlin2018bert}. Inspired by these advances, Transformer-based protein language models provide new opportunities for building global evolutionary models. A variety of Transformer-based models have been developed such as evolutionary scale modeling (ESM) \cite{rives2021biological, meier2021language}, ProGen \cite{madani2023large}, ProteinBERT \cite{brandes2022proteinbert}, Tranception \cite{notin2022tranception} and ESM-2 \cite{lin2023evolutionary}. 

\subsection{Hybrid approach via fine-tune pre-training}

Although global evolutionary models can learn a variety of sequences derived from natural evolution, they face challenges in concentrating on local information when predicting the effects of site-specific mutations in a target protein. To enhance the performance of global evolutionary models, fine-tuning strategies are subsequently implemented. Specifically, fine-tune strategy further refines the pre-trained global models with local information using MSAs or target training data. The fine-tuned eUniRep \cite{biswas2021low} shows significant improvement over UniRep \cite{alley2019unified}. Similar improvement was also reported for ESM models \cite{rives2021biological, meier2021language}. The Tranception model also proposed a hybrid approach combining a global autoregressive inference and a local retrieval inference from MSAs \cite{notin2022tranception}. Tranception achieved the advanced performance over other global and local models. 

With various language models proposed, comprehensive studies on various models and the strategy in building downstream model is necessary. A study explored different approaches utilizing the sequence embedding to build downstream models \cite{detlefsen2022learning}. Two other studies further benchmarked many unsupervised and supervised models in predicting protein fitness \cite{livesey2020using, hsu2022learning}.

\section{Structure-based topological data analysis (TDA) models}

Aided by advanced NLP algorithms, sequence-based models have become the dominant approach in MLPE \cite{yang2019machine, wittmann2021advances}. However, sequence-based models suffer from a lack of appropriate description of stereochemical information, such as cis-trans isomerism, conformational isomerism, enantiomers, etc. Therefore, sequence embeddings cannot distinguish stereoisomers, which are widely present in biological systems and play a crucial role in many chemical and biological processes. Structure-based models offer a solution to this problem. TDA has became a successful tool in building structure-based models for MLPE \cite{qiu2023persistent}. 

TDA is a mathematical framework based on algebraic topology \cite{edelsbrunner2008persistent, zomorodian2004computing}, which allows us to characterize complex geometric data, identify underlying geometric shapes, and uncover topological structures present in the data. TDA finds its applications in a wide range of fields, including neuroscience, biology, materials science, and computer vision. It is especially useful in situations where the data is complex, high-dimensional, and noisy, and where traditional statistical methods may not be effective. In this section, we provide an overview of various types of TDA methods (\autoref{table_tda}). In addition, we review graph neural networks, which are deep learning frameworks cognizant of topological structures, along with their applications in protein engineering. For those readers who are interested in the deep mathematical details of TDA, we have added a supplementary section dedicated to two TDA methods - persistent homology and persistent spectral graph (PSG) in Supplementary Methods.  

\begin{figure}[!htp]
	\centering
	\includegraphics[width=3.0in]{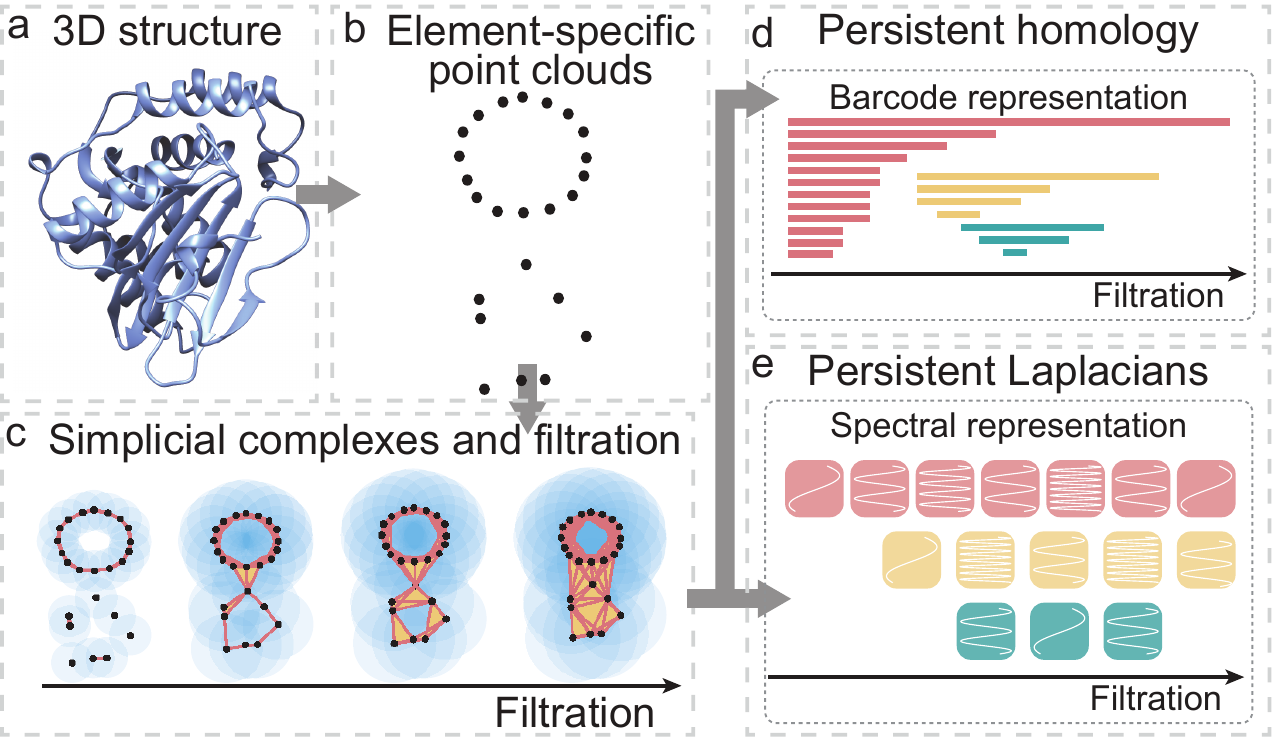}
	\captionsetup{width=14cm}
	\caption{\textbf{Conceptual illustration of the TDA-based protein modeling.}  (a). A three-dimensional protein structure. (b). Point cloud representation of protein structure. (c). Simplicial complexes and filtration provide multiscale topological representation of the point cloud. (d). Persistent homology characterizes topological evolution of the point cloud. (e). Persistent Laplacian characterizes shape evolution of the point cloud. }
\label{fig1}
\end{figure} 

{\small 
\begin{table*}[h]
    \centering
    \begin{tabular}{|c|c|c|c|}
        \hline
        \textbf{Method  } & \textbf{Topological space}  & \textbf{Node attribute } & \textbf{Edge attribute } \\
        \hline
        \multicolumn{4}{|c|}{\textbf{Homology-based}}\\ 
        \hline
        Persistent Homology \cite{edelsbrunner2008persistent, zomorodian2004computing}  &Simplicial complex & None&None\\
        \hline
        Element-specific PH (ESPH) \cite{cang2018integration}&Simplicial complex & Group labeled & Group labeled  \\
        \hline
        Persistent Cohomology \cite{cang2020persistent} &Simplicial complex & Labeled &  Labeled \\
        \hline
        Persistent Path Homology \cite{chowdhury2018persistent} &Path complex &Path &Directed \\ \hline
				Persistent Flag Homology \cite{lutgehetmann2020computing}& Flag complex& None & Directed \\  \hline
				Evolutionary homology \cite{cang2020evolutionary}& Simplicial complex & Weighted & Weighted \\ \hline
				Weighted persistent homology \cite{meng2020weighted}& Simplicial complex & Weighted & Weighted \\
        \hline
        \multicolumn{4}{|c|}{\textbf{Laplacian-based}}\\ 
        \hline
				 Persistent Spectral Graph \cite{wang2020persistent,memoli2022persistent} &Simplicial complex  &None &None \\ 
				 \hline    
        Persistent Hodge Laplacians \cite{chen2021evolutionary} &Manifold &Continuum&Continuum\\
        \hline
        Persistent Sheaf Laplacians \cite{wei2021persistent} &Cellular complex& Labeled & Sheaf relation\\
        \hline
        Persistent Path Laplacians \cite{wang2023persistent} &Path complex  &Path &Direction \\
        \hline
        Persistent Hypergraph \cite{liu2021persistent}&Hypergraph&Hypernode &Hyperedge \\
        \hline
        Persistent Directed Hypergraphs \cite{chen2023persistent} &Hypergraph&Hypernode &Directed hyperedge \\
       	\hline
    \end{tabular}
        \caption{Summary of topological data analysis (TDA) methods for structures.}
        \label{table_tda}
\end{table*}
}
 \subsection{Homology }\label{sec_TDA} 

The basic idea behind TDA is to represent the data as a point cloud in a high-dimensional topological space, and then study the topological invariants of this space, such as the genus number, Betti number, and Euler characteristic. Among them, the Betti numbers, specifically Betti zero, Betti one, and Betti two, can be interpreted as representing connectedness, holes, and voids, respectively \cite{kaczynski2004computational, wasserman2018topological}. These numbers can be computed as the ranks of the corresponding homology groups in appropriate dimensions.

Homology groups are algebraic structures that are associated with topological spaces \cite{kaczynski2004computational}. They provide information about the topological connectivity of geometric objects. The basic idea behind homology is to consider the cycles and boundaries of a space. Loosely speaking, a cycle is a set of points in the space that form a closed loop, while a boundary is a set of points that form the boundary of some region in the space.
The homology group of a space is defined as the group of cycles modulo the group of boundaries. That is, we identify two cycles that differ by a boundary and consider them to be equivalent. The resulting homology group encodes information about the Betti numbers of the space.

Homology theory has many applications in mathematics and science. It is used to classify topological spaces in category theory, to study the properties of manifolds in differential geometry and algebraic geometry, and to analyze data in various scientific fields \cite{kaczynski2004computational}. However, the original homology groups offer truly geometry-free representations and are too abstract to carry sufficient geometric information of data.
Persistent homology was designed to improve homology groups' ability for data analysis.

\subsection{Persistent homology}

Persistent homology is a relatively new tool in algebraic topology that is designed to incorporate multiscale topological analysis of data \cite{edelsbrunner2008persistent, zomorodian2004computing}. The basic idea behind persistent homology is to construct a family of geometric shapes of the original data by filtration (\autoref{fig1}c). Filtration systematically enlarges the radius of each data point in a point cloud, leading to a family of topological spaces with distinct topological dimensions and connectivity. Homology groups are built from the family of shapes, giving rise to systematic changes in topological invariants, or Betti numbers, at various topological dimensions and geometric scales. Topological invariants based on Betti numbers are expressed in terms of persistence barcodes \cite{ghrist2008barcodes} (\autoref{fig1}d), persistence diagrams \cite{cohen2005stability}, persistence landscapes \cite{bubenik2015statistical}, or persistence images \cite{adams2017persistence}. Persistent topological representations are widely used in applications, particularly in association with machine learning models \cite{cang2015topological}.
 
Persistent homology is the most important approach in TDA (see Table \ref{table_tda} for a summary of major TDA approaches). It reveals the shape of data in terms of the topological invariants and has had tremendous success in scientific applications, including image and signal processing \cite{clough2020topological}, machine learning \cite{pun2018persistent}, biology \cite{cang2015topological}, and neuroscience\cite{stolz2017persistent}. Nonetheless, to effectively analyze complex biomolecular data, persistent homology requires further refinement and adjustment. \cite{wei2023topological}.

\subsection{Persistent cohomology and element-specific persistent homology}

One major limitation of persistent homology is that it fails to describe heterogeneous information of data point \cite{cang2020persistent}. In other words, it treats all entries in the point cloud equally without considering other important information about the data. Biomolecules, for example, contain many different element types and each atom may have a different atomic partial charge, atomic interaction environment, and electrostatic potential function that cannot be captured by persistent homology. Thus, it is crucial to have a topological technique that can incorporate both geometric and nongeometric information into a unified framework.

Persistent cohomology was developed to provide such a mathematical paradigm \cite{cang2020persistent}. In this framework, nongeometric information can either be prescribed globally or reside locally on atoms, bonds, or many-body interactions. In topological terminology, nongeometric information is defined on simplicial complexes. This persistent cohomology-based approach can capture multiscale geometric features and reveal non-geometric interaction patterns through topological invariants, or enriched persistence barcodes. It has been demonstrated that persistent cohomology outperforms other methods in benchmark protein-ligand binding affinity prediction datasets \cite{cang2020persistent}, which is a non-trivial problem in computational drug discovery. 
 
An alternative approach for addressing the limitation of persistent homology is to use element-specific persistent homology (ESPH) \cite{cang2018integration}. The motivation behind ESPH is the same as that for persistent cohomology, but ESPH is relatively simple. Basically, atoms in the original biomolecule are grouped according to their element types, such as C, N, O, S, H, etc. Then, their combinations, such as CC, CN, CO, etc., are identified, and persistent homology analysis is applied to the atoms in each element combination, resulting in ESPH analysis. As a result, ESPH reduces geometric and biological complexities and embeds chemical and biological information into topological abstraction. The ESPH approach was used to win the D3R Grand Challenges, a worldwide competition series in computer-aided drug design \cite{nguyen2019mathematical}.

\subsection{Persistent topological Laplacians}

However, aforementioned TDA methods are still limited in describing complex data, such as its lack of description of non-topological changes (i.e., homotopic shape evolution) \cite{qiu2023persistent}, its incapability of coping with directed networks and digraphs (i.e., atomic partial charges and polarizations, gene regulation networks), and its inability to characterize structured data (e.g., functional groups, binding domains, and motifs) \cite{wei2023topological}. These limitations necessitate the development of innovative strategies.

Persistent topological Laplacians (PTLs) are a new class of mathematical tools designed to overcome the aforementioned challenges in TDA \cite{wei2023topological}. One of the first methods in this class is the PSG \cite{wang2020persistent}, also known as persistent  combinatorial  Laplacians  \cite{wang2020persistent} or persistent Laplacians
\cite{memoli2022persistent}. PSGs have both harmonic spectra with zero eigenvalues and non-harmonic spectra with non-zero eigenvalues (\autoref{fig1}e). The harmonic spectra recover all the topological invariants from persistent homology, while the non-harmonic spectra capture the homotopic shape evolution of data that cannot be described by persistent homology \cite{wei2023topological}. PSGs have been used for accurate forecasting of emerging dominant SARS-CoV-2 variants BA.4/BA.5 \cite{chen2022persistent}, facilitating machine learning-assisted protein engineering predictions \cite{qiu2023persistent}, and other applications \cite{meng2021persistent}.

Like persistent homology, persistent Laplacians are limited in their ability to handle directed networks and atomic polarizations. To address these limitations, persistent path Laplacians have been developed \cite{wang2023persistent}. Their harmonic spectra recover the topological invariants of persistent path homology \cite{chowdhury2018persistent}, while their non-harmonic spectra capture homotopic shape evolution. Both persistent path Laplacians and persistent path homology were developed as a generalization of the path complex \cite{grigor2020path}.

None of the PTLs mentioned above are capable of handling different types of elements in a molecule as persistent cohomology does. To overcome this limitation, persistent sheaf Laplacians \cite{wei2021persistent} were designed, inspired by persistent cohomology \cite{cang2020persistent}, persistent Laplacians \cite{wang2020persistent}, and sheaf Laplacians for cellular sheaves \cite{hansen2019toward}. The aim of persistent sheaf Laplacians is to discriminate between different objects in a point cloud. By associating a set of non-trivial labels with each point in a point cloud, a persistent module of sheaf cochain complexes is created, and the spectra of persistent sheaf Laplacians encode both geometrical and non-geometrical information \cite{wei2021persistent}. The theory of persistent sheaf Laplacians is an elegant method for the fusion of different types of data and opens the door to future developments in TDA,   geometric data analysis, and algebraic data analysis.

Persistent hypergraph Laplacians enable the topological description of internal structures or organizations in data \cite{liu2021persistent}. Persistent hyperdigraph Laplacians further allow for the topological Laplacian modeling of directed hypergraphs \cite{chen2023persistent}. These persistent topological Laplacians can be utilized to describe intermolecular and intramolecular interactions. As protein structures are inherently multiscale, it is natural to apply persistent hypergraph Laplacians and persistent hyperdigraph Laplacians to delineate the protein structure-function relationship.  

Finally,  unlike all the aforementioned PTLs,  evolutionary de Rham-Hodge Laplacians or persistent Hodge Laplacians are defined on a family of filtration-induced differentiable manifolds \cite{chen2021evolutionary}. They are particularly valuable for the  multiscale topological analysis of volumetric data. Technically, a similar algebraic topology structure is shared by persistent Hodge Laplacians and persistent Laplacians, but the former is a continuum theory for volumetric data and the latter is a discrete formulation for point cloud. As such,  their underlying mathematical definitions, i.e., differential forms on manifolds and simplicial complexes on graphs, are sharply different.

\subsection{Deep graph neural networks and topological deep learning}

Similar to topological data analysis, graph- and topology-based deep learning models have been proposed to capture connectivity and shape information of protein structure data. Graph neural networks (GNNs) consider the low-order interactions between vertices by aggregating information from neighbor vertices. A variety of popular graph neural network layers has been proposed, such as convolution graph networks (GCN) \cite{kipf2016semi}, graph attention networks (GAT) \cite{velivckovic2017graph}, graph sample and aggregate (GraphSAGE) \cite{hamilton2017inductive},  Graph Isomorphism Network (GIN) \cite{xu2018powerful}, and gated graph neural network \cite{li2015gated}. 

With variety of architectures of GNN layers, self-supervised learning models are widely used for representation learning of graph-based data. Graph autoencoder (GAE) and variational graph autoencoder (VGAE) consist of both encoder and decoder, where the decoder employ a linear inner product to reconstruct adjacent matrix \cite{kipf2016variational}. While most of graph-based self-supervised models only have encoder. Deep graph infomax (DGI) maximizes mutual information between a graph's local and global features to achieve self-supervised learning \cite{velivckovic2018deep}. Graph contrastive learning (GRACE) constructs positive and negative pairs from a single graph, and trains a GNN to differentiate between them \cite{you2020graph}. Self-supervised graph transformer (SSGT) uses masked node prediction to train the model. Given a masked graph, it tries to predict the masked node's attributes from the unmasked nodes \cite{rong2020self}. 

In applications to learning protein structures, GCNs have been widely applied to building structure-to-function map of proteins \cite{li2021structure, gligorijevic2021structure}. Moreover, self-supervised models provide powerful pre-trained model in learning representation of protein structures. GeoPPI \cite{liu2021deep} proposed a graph neural network-based autoencoder to extract structural embedding at the protein-protein binding interface. The subsequent downstream models allow accurate predictions for protein-protein binding affinity upon mutations \cite{liu2021deep} and further design effective antibody against SARS-CoV-2 variants \cite{shan2022deep}. GRACE has been applied to learn geometric representation of protein structures \cite{zhang2022protein}. To adopt the critical biophysical properties and interactions between residues and atoms in protein structures, graph-based self-supervised learning models have been customized to achieve the specific functions. The inverse protein folding protocol was proposed to capture the complex structural dependencies between residues in its representation learning \cite{ingraham2019generative, hsu2022esm}. OAGNNs was proposed to better sense the geometric characteristics such as nner-residue torsion angles, inter-residue orientations in its representation learning \cite{li2022orientation}.

 Topological deep learning,  proposed by Cang and Wei in 2017 \cite{cang2017topologynet}, is an emerging paradigm.   It integrates topological representations with deep neural networks   for protein fitness learning and prediction \cite{cang2017topologynet, nguyen2019mathematical,qiu2023persistent}. Similar graph and topology-based deep learning architectures have also been proposed to capture connectivity and shape information of protein structure data \cite{chen2022persistent,chen2023persistent}. 
Inspired by TDA,  high-order interactions among neural nodes  were proposed in $k$-GNNs \cite{morris2019weisfeiler} and simplicial neural networks \cite{ebli2020simplicial}.

\section{Artificial intelligence-aided protein engineering}

Protein engineering is a typical black-box optimization problem, which focuses on finding the optimal solution without explicitly knowing the objective function and its gradient. In protein engineering, the goal in designing algorithms for this problem is to efficiently search for the best sequence within a large search space:

\begin{equation}
\label{eq_box}
x^{*}=\argmax_{x\in \mathcal{S}}f(x),
\end{equation}
where $\mathcal{S}$ is an unlabeled candidate sequence library, $x$ is a sequence in the library and $f(x)$ is the unknown sequence-to-fitness map for optimization. The fitness landscape, $f(\mathcal{S})$, is a high-dimensional surface that maps amino acid sequences to properties such as activity, selectivity, stability, and other physicochemical features.  

There are two practical challenges in protein engineering. First, the fitness landscape is usually epistatic \cite{wu2016adaptation, podgornaia2015pervasive}, where the contribution of individual amino acid residues to protein fitness have dependency to each other. The interdependence leads to complex, non-linear interactions among different residues. In other word, the fitness landscape contains large number of local optima. For example, in a four-site mutational fitness landscape for GB1 protein with $20^4=160,000$ mutations, 30 local maximum fitness peaks were found \cite{wu2016adaptation}. Either traditional directed evolution experiments such as single-mutation walk and recombination, or machine learning models, is difficult to find the global optima without trapped at local one. Second, protein engineering process usually collects limited number of data comparing to the huge sequence library. There are an enormous number of ways to mutate any given protein: for a 300-amino-acid protein, there are 5,700 possible single-amino-acid substitutions and 32,381,700 ways to make just two substitutions with the 20 canonical amino acids \cite{yang2019machine}. Even with high-throughput experiments, only a small fraction of the sequence library can be screened. Despite this, many systems only have low-throughput assays such as membrane proteins \cite{zhang2023structural}, making the process more difficult. 

With enriched data-driven protein modeling approaches from protein sequences to structures, recent advanced machine learning methods have been widely developed to accelerate protein engineering in silico (\autoref{fig2}a) \cite{narayanan2021machine, wittmann2021advances, yang2019machine, freschlin2022machine, hie2022adaptive}. Utilizing a limited experimental capacity, machine learning models can effectively augment the fitness evaluation process, enabling the exploration of a vast search space $\mathcal{S}$. This approach facilitates the discovery of optimal solutions within complex design spaces, despite constraints on the number of trials or experiments.

Using a limited number of experimentally labeled sequences, machine learning models can carry out zero-shot or few-shot predictions \cite{wittmann2021advances}. The accuracy of these predictions largely depends on the distribution of the training data, which influences the model's ability to generalize to new sequences. Concretely, if the training data is representative or closer to a given sequence, the model is more likely to make accurate predictions for that specific sequence. Conversely, if the training data is not representative or distant from the given sequence, the model's predictive accuracy may be compromised, leading to less reliable results. Therefore, MLPE are usually an iterative process between machine learning models and experimental screens. Incorporating the exploration-exploitation trade-off in this context is essential for achieving optimal results. During the iterative process, the model must balance exploration, where it seeks uncertain regions that machine learning models have low accuracy, with exploitation, where it refines and maximizes fitness based on previously gained knowledge.  A right balance is critical to preventing overemphasis on either exploration or exploitation leading, which may lead to suboptimal solutions. In particular, the epistatic nature of protein fitness landscapes influences the exploration-exploitation trade-off in the design process.

MLPE methods need to take the experimental capacity into account when attempt to balance the exploitation-exploration. In this section, we discuss different strategies upon the number of experimental capacity. First, we discuss zero-shot strategy when no labeled experimental data is available. Second, we discuss supervised models for performing greedy search (i.e., exploitation). Last, we discuss uncertainty quantification models that balance exploration and exploitation trade-off.

\subsection{Unsupervised zero-shot strategy}

First, we review the zero-shot strategy that interrogates protein fitness with an unsupervised manner (\autoref{fig2}b and \autoref{table_comp}). This is designed for the scenarios in the early stage designs where no experiments have been conducted or the experimentally labeled data is too limited allowing accurate fitness predictions from supervised models \cite{wittmann2021advances, qiu2023persistent}. They delineate a fitness landscape at the early stage of protein engineering. Essential residues can be identified and prioritized for mutational experiments, allowing for a more targeted approach to protein engineering \cite{riesselman2018deep}. Additionally, the initial fitness landscape can be utilized to filter out protein candidates with a low likelihood of exhibiting the desired functionality.  By focusing on sequences with higher probabilities, protein engineering process can be made more efficient and effective \cite{wittmann2021informed}.

Zero-shot predictions rely on the model's ability to recognize patterns in naturally observed proteins, enabling it to make informed predictions for new sequences without having direct training data for the target protein. As discussed in Section \ref{sec_PLM}, protein language models, particularly generative models, learn the distribution of naturally observed proteins which are usually functional. The learned distribution can be used to assess the likelihood that a newly designed protein lies within the distribution of naturally occurring proteins, thus providing valuable insights into its potential functionality and stability \cite{wittmann2021advances}. 

VAEs are popular local evolutionary models for zero-shot predictions such as DeepSequence \cite{riesselman2018deep} and EVE models \cite{frazer2021disease}. In VAEs, the conditional probability distribution $p(x\vert z, \theta)$ is the decoder in a form of neural network with parameters $\theta$, where $x$ is the sequence being query and $z$ is its latent space variable. Similar, encoder, $q(z\vert x,\phi)$, is modeled by another neural network with parameters $\phi$ to approximate the true posterior distribution $p(z\vert x)$. For a given sequence $x$, its probabilistic likelihood in VAEs is $p(x\vert \theta)$ parameterized by parameters $\theta$. Direct computation of this probability, $p(x\vert \theta)=\int p(x\vert z,\theta)dz$, is intractable in the general case.  The evidence lower bound (ELBO) forming a variation inference \cite{kingma2013auto} provides a lower bound of the log likelihood:
\begin{equation}
\log p(x\vert\theta)\geq {\rm ELBO}(x)=\mathbb{E}_q\log p(x\vert z, \theta)-{\rm KL}\left(q(z\vert x,\phi)\| p(z)\right).
\end{equation}
ELBO is taken as the scoring function to quantify the mutational likelihood of each query sequence. The ELBO-based zero-shot predictions show advanced performance reported in multiple works \cite{hsu2022learning,livesey2020using,qiu2023persistent}. 

Transformer is the currently state-of-the-art model which has been used in many supervised tasks \cite{rives2021biological}. It learns a global distribution of nature proteins. It has also been proved to have advanced performance for zero-shot predictions \cite{hsu2022learning, meier2021language}. The training of Transformer uses mask filling that refers to the process of predicting masked amino acid in a given input sequence by leveraging the contextual information encoded in the Transformer's self-attention mechanism \cite{vaswani2017attention, devlin2018bert}. The mask filling procedure creates a classification layer on the top of the Transformer architecture. Given a sequence $x$, the masked filling classifier generate probability distributions for amino acids at masked positions. Suppose $x$ has $L$ amino acids $x=x_1x_2\cdots x_L$, by masking a single amino acid at $i$-th position, the classifier calculates the conditional probability of $p(x_i\vert x^{(-i)})$, where $x^{(-i)}$ is the remaining sequence excluding the masked $i$-th position. To reduce the computational cost, the pseudo-log-likelihoods (PLLs) are usually used to estimate the log-likelihood of a given sequence \cite{hsu2022learning, wittmann2021informed}:
\begin{equation}
\mbox{PLL}(s)=\sum_{i=1}^L\log P(s_i\vert s^{(-i)}).
\end{equation}
The PPLs assume the independence between amino acids. To consider the dependence between amino acids, one can calculate the conditional probability by summing up all possible factorization \cite{wittmann2021informed}. But this approach leads to much higher computational cost.

Furthermore, many different strategies have been employed to make zero-shot predictions. Fine-tune model can improve the predictions by combining both local and global evolutionary models \cite{biswas2021low}. Tranception scores combine global autoregressive inference and an local MSAs retrieval inference to make more accurate predictions. In addition to these sequence-based models, the structure-based GNN-based models including ESM-1F \cite{hsu2022esm} and RGC \cite{tian2023sequence} have also been proposed by utilizing large-scale structural data from AlphaFold2. However, the structure-based model is still limited in accuracy comparing to sequence-based models.

\subsection{Supervised regression models}
Supervised regression models are among the most prevalent approaches used in guiding protein engineering, as they enable greedy search strategies to maximize protein fitness (\autoref{fig2}c). These models, including statistical, machine learning, and deep learning techniques, rely on a set of labeled data as their training set to predict the fitness landscape. By leveraging the information contained within the training data, supervised regression models can effectively estimate the relationship between protein sequences and their fitness, providing valuable insights for protein engineering and optimization \cite{yang2019machine, narayanan2021machine}.

A variety of supervised models have been applied to predict protein fitness. In general, statistical models and machine learning models such as linear regression \cite{fox2007improving}, ridge regression \cite{hsu2022learning}, support vector machine (SVM) \cite{guo2008using}, random forest \cite{zhang2020mutabind2}, gradient boosting tree \cite{cang2017analysis} have accurate performance for small training set. And deep learning methods such as deep neural networks \cite{aghazadeh2021epistatic}, convolutional neural networks (CNNs) \cite{wang2020topology}, attention-based neural networks \cite{dallago2021flip} are more accurate with large size of training data. However, in protein engineering, the size of training data increases sequentially which make the supervised models difficult to provide accurate performance all time. Alternatively, the ensemble regression was proposed to provide robust fitness predictions despite of training data size \cite{wittmann2021advances, bryant2021deep}. The ensemble regression average predictions from multiple supervised models and they provide more accurate and robust performance than single model \cite{qiu2023persistent}. To remove the inaccurate models in the average, cross-validation is usually used to rank accuracy of each model and only top models are taken to average the predictions. Paired with the zero-shot strategy, the ensemble regression trained on informed training set pre-selected by zero-shot predictions can efficiently pick up the global optimal protein with a few round of experiments \cite{wittmann2021informed, qiu2021cluster, qiu2022clade}. And such approach has been applied to enable resource-efficient engineering CRISPR-Cas9 genome editor activities \cite{thean2022machine}.

Rather than the architectures of supervised models, the predictive accuracy highly rely on the amount of information obtained from the featurization process (\autoref{table_comp}). The physical-chemical properties extract the properties of individual amino acids or atoms \cite{georgiev2009interpretable}. The energy-based scores provide descriptions for the overall property of the target protein \cite{schymkowitz2005foldx}. However, neither of them successfully take the complex interactions between residues and atoms into account. To tackle this challenge, recent mathematics-initiated topological and geometric descriptors achieved great success in predicting protein fitness including protein-protein interactions \cite{wang2020topology}, protein stability \cite{cang2017analysis}, enzyme activity, and antibody effectivity \cite{qiu2023persistent}. The aforementioned descriptors (Section \ref{sec_TDA}) extract structural information from atoms at different characteristic lengths. Furthermore, the sequence-based protein language models provide another featurization strategies. The deep pre-trained models have the latent space which provide the informative representation of each given sequence. Building supervised models from the deep embedding exhibits accurate performance \cite{qiu2023persistent, shen2022svsbi}. Recent works combine different types of sequence-based features \cite{luo2021ecnet, hsu2022learning} or combine structure-based and sequence-based features \cite{qiu2023persistent} show the complementary roles of different featurization approaches. 

{\small 
\begin{table}[]
\centering
\begin{tabular}{|l|r|r|r|r|}
\hline
\multicolumn{5}{|c|}{\textbf{Zero-shot predictors}}\\
\hline
\multirow{2}{*}{Model name} & \multicolumn{4}{|c|}{training set size} \\
\cline{2-5}
&\multicolumn{4}{|c|}{$0$} \\
\hline
ESM-1b PLL \cite{rives2021biological, hsu2022learning} & \multicolumn{4}{|c|}{0.435} \\
\hline
eUniRep PLL \cite{georgiev2009interpretable} & \multicolumn{4}{|c|}{0.411} \\
\hline
EVE  \cite{frazer2021disease} & \multicolumn{4}{|c|}{0.497} \\
\hline
Tranception \cite{notin2022tranception} & \multicolumn{4}{|c|}{0.478} \\
\hline
DeepSequence \cite{riesselman2018deep} & \multicolumn{4}{|c|}{\textbf{0.504}} \\
\hline
\multicolumn{5}{|c|}{\textbf{Supervised models}}\\
\hline
\multirow{2}{*}{Embedding name} & \multicolumn{4}{|c|}{training set size} \\
\cline{2-5}
& $24$ & $96$ & $168$ & $240$ \\
\hline
Persistent Homology \cite{qiu2023persistent} & 0.263 & 0.432 & 0.496 & 0.534 \\
\hline
Persistent Laplacian  \cite{qiu2023persistent} & \textbf{0.280} & \textbf{0.457} & \textbf{0.525} & \textbf{0.564} \\
\hline
ESM-1b \cite{rives2021biological} & 0.219 & 0.421 & 0.494 & 0.537 \\
\hline
eUniRep \cite{biswas2021low} & 0.259 & 0.432 & 0.485 & 0.515 \\
\hline
Georgiev \cite{georgiev2009interpretable} & 0.169 & 0.326 & 0.402 & 0.446 \\
\hline
UniRep \cite{alley2019unified} & 0.183 & 0.347 & 0.420 & 0.462 \\
\hline
Onehot & 0.132 & 0.317 & 0.400 & 0.450 \\
\hline
Bepler \cite{bepler2018learning}& 0.139 & 0.287 & 0.353 & 0.396 \\
\hline
TAPE LSTM \cite{rao2019evaluating} & 0.259 & 0.436 & 0.492 & 0.522 \\
\hline
TAPE ResNet \cite{rao2019evaluating} & 0.080 & 0.216 & 0.305 & 0.358 \\
\hline
TAPE Transformer \cite{rao2019evaluating} & 0.146 & 0.304 & 0.371 & 0.418 \\
\hline
\end{tabular}
\caption{\textbf{Comparisons for fitness predictors. } 
Results were adopted from TopFit  \cite{qiu2023persistent}. Performance was reported by average Spearman correlation over 34 DMS datasets and 20 repeats. Supervised model use ensemble regression from 18 regression models \cite{qiu2023persistent}. }
\label{table_comp}
\end{table}
}
\subsection{Active learning models for exploration-exploitation balance}

With the extensive accurate protein-to-fitness machine learning models, active learning further designs iterative strategy between models and experiments to sequentially optimize fitness with the consideration of exploitation-exploration trade-off (\autoref{fig2}d-e) \cite{hie2022adaptive}. 

To balance the exploitation-exploration trade-off, the supervised models require to predict not only the protein fitness but also quantify the uncertainty of the given protein \cite{greenman2022benchmarking}. The most popular uncertainty quantification in protein engineering is Gaussian process (GP)  \cite{rasmussen2003gaussian}, which automatically calibrate the balance. Especially, GP using the upper confidence bounds (UCBs) acquisition has efficient convergent rate theoretically for solving the black-box optimization (\autoref{eq_box}). A variety protein engineering employed GP to accelerate the fitness optimization. For examples, the light-gated channelrhodopsins (ChRs) were engineered to improve photocurrence and light sensitivity \cite{bedbrook2019machine, bedbrook2017machine}, green fluorescent protein has been engineered to become yellow fluorescence \cite{saito2018machine}, acyl-ACP reductase was engineered to improve fatty alcohol production \cite{greenhalgh2021machine}, P450 enzyme has been engineered to improve thermostability \cite{romero2013navigating}.

The tree-based search strategy is also efficient by building a hierarchical search path, such as the hierarchical optimistic optimization (HOO) \cite{bubeck2011x}, the deterministic optimistic optimization (DOO), and the simultaneous optimistic optimization (SOO) \cite{munos2011optimistic}. To handle the discrete mutational space in protein engineering, an unsupervised clustering approach was employed to construct the hierarchical tree structure \cite{qiu2021cluster,qiu2022clade}. 

Recently, researchers have turned to generative models to quantify uncertainty in protein engineering, employing methods such as Variational Autoencoders (VAEs) \cite{kingma2013auto, riesselman2018deep, frazer2021disease}, generative adversarial networks (GANs) \cite{creswell2018generative,gupta2019feedback}, and autoregressive language models \cite{shin2021protein, notin2022tranception}. Generative models are a class of machine learning algorithms that aim to learn the underlying data distribution of a given dataset, in order to generate new, previously unseen data points that resemble the training data. These models capture the inherent structure and patterns present in the data, enabling them to create realistic and diverse samples that share the same characteristics as the original data. For examples, ProGen \cite{madani2023large} is a large language model that generate functional protein sequences across diverse families. A Transformer-based antibody language models utilize fine-tuning processes to assist design antibody \cite{bachas2022antibody}. Recently, a novel Transformer-based model called ReLSO has been introduced \cite{castro2022transformer}. This innovative approach simultaneously generates protein sequences and predicts their fitness using its latent space representation. The attention-based relationships learned by the jointly trained ReLSO model offer valuable insights into sequence-level fitness attribution, opening up new avenues for optimizing proteins. 



\section{Conclusions and future directions}
In this review, we  discuss the advanced deep protein language models for protein modeling.
We further provide an introduction of topological data analysis methods and their applications in protein modeling.  Relying on both structure-based and sequence-based models, MLPE methods were widely developed to accelerate protein engineering. In the future, various machine learning and deep learning will have potential perspectives in protein engineering. 

\subsection{Accurate structure prediction methods enhanced accurate structure-based models}

Comparing to sequence data, three-dimensional protein structural data offer more comprehensive and explicit descriptions of the biophysical properties of a protein and its fitness. As a result, structure-based models usually provide superb performance than sequence-based models for supervised tasks with small training set \cite{qiu2023persistent, cang2017analysis}.

As protein sequence  databases continue to grow, self-supervised models demonstrate their ability to effectively model proteins using large-scale data. The protein sequence database provides a vast amount of resources for building sequence-based models, such as UniProt \cite{uniprot2021uniprot} database contains hundreds of millions sequences. In contrast, protein structure databases are comparatively limited in size. The largest among them, Protein Data Bank (PDB), contains only 205 thousands of protein structures as of 2023 \cite{berman2000protein}. Due to the abundance of data resources, sequence-based models typically outperform structure-based models significantly \cite{tian2023sequence}.

To address the limited availability of structure data, researchers have focused on developing highly accurate deep learning techniques aimed at enabling large-scale structure predictions. These state-of-the-art methodologies have the potential to significantly expand the database of known protein structures. Two prominent methods are AlphaFold2 \cite{jumper2021highly} and RosettaFold \cite{baek2021accurate}, which have demonstrated remarkable capabilities in predicting protein structures with atomic-level accuracy. By harnessing the power of cutting-edge deep learning algorithms, these tools have successfully facilitated the accurate prediction of protein structures, thus contributing to the expansion of the structural database.

Both AlphaFold2 and RosettaFold are alignment-based, which rely on MSAs of the target protein for structure prediction. Alignment-based approaches can be highly accurate when there are sufficient number of homologous sequences (that is, MSAs depth) in the database. Therefore, these methods may have reduced accuracy with low MSAs depth in database. In addition, the MSAs search is time consuming which slows down the prediction speed. Alternatively, alignment-free methods have also been proposed to tackle these limitations \cite{kandathil2023machine}. An early work RGN2 \cite{chowdhury2022single} exhibits more accurate predictions than AlphaFold2 on orphans proteins which lack of MSAs. 
Supervised transformer protein language models predict orphan protein structures   \cite{wang2022single}. With the development of variety of large-scale protein language models in recent years, the alignment-free structural prediction methods incorporate with these models to exhibit their accuracy and efficiency. For example, ESMFold \cite{lin2023evolutionary} and OmegaFold \cite{wu2022high} achieve similar accuracy with AlphaFold2 with faster speed. Moreover, extensive language model-based methods were developed for structural predictions of single-sequence and orphan proteins \cite{fang2022helixfold,barrett2022so,wu2022tfold,weissenow2022ultra}. Large-scale protein language models will provide powerful toolkit for protein structural predictions. 

In building protein fitness model, the structural TDA-based model has exemplified that the AlphaFold2 structure is as reliable as the experimental structure \cite{qiu2023persistent}. The zero-shot model, ESM-IF1, also shows advanced performance with coupling with the large structure AlphaFold database \cite{hsu2022esm}. In the light of the revolutionary structure predictive models, structure-based models will open up a new avenue in protein engineering, from directed evolution to de novo design \cite{bordin2022novel,chidyausiku2022novo}. More sophisticated TDA methods will be demanded to handle the large-scale datasets. Large-scale deep graph neural networks will need to be further developed, for example, to consider the high-order   interactions using simplicial neural networks \cite{ebli2020simplicial, keros2022dist2cycle}.

\subsection{Large highthroughput datasets enabled larger scale models}

Current MLPE methods are usually designed for limited training set. The ensemble regression is an effective approach to accurately learn the fitness landscape with small but increasing size of training sets from deep mutational scanning  \cite{wittmann2021informed}.

The breakthrough biotechnology, next-generation sequencing (NGS) \cite{schuster2008next} largely enhances the capacity of DMS for collecting supervised fitness data in various protein systems \cite{podgornaia2015pervasive,wu2016adaptation, sarkisyan2016local}. The resulting large-scale deep mutational scanning  databases expand the exploration range of protein engineering. Deeper machine learning models are emerging to enhance the accuracy and adaptivity for protein engineering.

\section{Competing interests}
No competing interest is declared.

\section{Author contributions statement}
Y.Q. and G.W.W conceived, wrote, and revised the manuscript.

\section{Acknowledgments}
This work was supported in part by NIH grants  R01GM126189 and  R01AI164266, NSF grants DMS-2052983,  DMS-1761320, and IIS-1900473,  NASA grant 80NSSC21M0023,  Michigan Economic Development Corporation, MSU Foundation,  Bristol-Myers Squibb 65109, and Pfizer. 

\clearpage
\bibliographystyle{unsrt}

\clearpage
\subfile{supplement}

\end{document}

%% file: supplement.tex
    \maketitle




\section{Mathematical theory of topological data analysis (TDA)}
\label{sec_TDA}

\subsection{Simplicial complex and chain complex}
Graph is a representation for a point cloud consisting of vertices and edges for modeling pairwise interactions, such as atoms and bonds in molecules. Simplicial complex, the generalization of graph, constructs more enriched shapes to include high dimensional objects. A simplicial complex is composed of simplexes up to certain dimensions. A $k$-simplex, $\sigma^k$, is a convex hull of $k+1$ affinely independent points $v_0,\ v_1,\ v_2,\ \cdots,\ v_k$:
\begin{equation}
\label{eq1}
\sigma^k:=\left[v_0,\ v_1,\ v_2,\ \cdots,\ v_k\right]=\left\{\sum_{i=0}^{k}\lambda_iv_i\middle\vert \sum_{i=0}^{k}\lambda_i=1; \lambda_i\in [0,1], \ \forall i \right\}.
\end{equation} 
In Euclidean space, 0-simplex is a point, 1-simplex is an edge, 2-simplex is a triangle, and 3-simplex is a tetrahedron. The $k$-simplex can describe abstract simplex for $k>3$. 

A subset of the $k+1$ vertices of a $k$-simplex, $\sigma^k$, with $m+1$ vertices forming a convex hull in a lower dimension and is called an $m$-face of the $k$-simplex $\sigma^m$, denoted as $\sigma^m\subset\sigma^k$. A simplicial complex $K$ is a finite collection of simplexes satisfying two conditions: 

1) Any face of a simplex in $K$ is also in $K$.

2) The intersection of any two simplexes in $K$ is either empty or a shared face. 

The interactions between two simplexes can be described by adjacency. For example, in graph theory, two vertices (0-simplexes) are adjacent if they share a common edge (1-simplex). Adjacency for $k$-simplexes with $k>0$ includes both upper and lower adjacency. Two distinct $k$-simplexes, $\sigma_1$ and $\sigma_2$, in $K$ are upper adjacent, denoted $\sigma_1\sim_U\sigma_2$, if both are faces of a $(k+1)$-simplex in $K$, called a common upper simplex. Two distinct $k$-simplexes, $\sigma_1$ and $\sigma_2$, in $K$ are lower adjacent, denoted $\sigma_1\sim_L\sigma_2$, if they share a common $(k-1)$-simplex as their face, called a common lower simplex. Either common upper simplex or common lower simplex is unique for two upper or lower adjacent simplexes. The upper degree of a $k$-simplex, $\mbox{deg}_U(\sigma^k)$, is the number of $(k+1)$-simplexes in $K$ of which $\sigma^k$ is a face; the lower degree of a $k$-simplex, $\mbox{deg}_L(\sigma^k)$, is the number of nonempty $(k-1)$-simplexes in $K$ that are faces of $\sigma^k$, which is always $k+1$. The degree of $k$-simplex ($k>0$) is defined as the sum of its upper and lower degree 
\begin{equation}
\mbox{deg}(\sigma^k)=\mbox{deg}_U(\sigma^k)+\mbox{deg}_L(\sigma^k)=\mbox{deg}_U(\sigma^k)+k+1.
\end{equation}
For $k=0$, the degree of a vertex is:
\begin{equation}
\mbox{deg}(\sigma^0)=\mbox{deg}_U(\sigma^0).
\end{equation}
A simplex has orientation determined by the ordering of its vertices, except 0-simplex. For example, clockwise and anticlockwise orderings of three vertices determine the two orientation of a triangle. Two simplexes, $\sigma_1$ and $\sigma_2$, defined on the same vertices are similarly oriented if their orderings of vertices differ from an even number of permutations, otherwise, they are dissimilarly oriented. 

Algebraic topology provides a tool to calculate simplicial complex. A $k$-chain is a formal sum of oriented $k$-simplexes in $K$ with coefficients on $\mathbb{Z}$. The set of all $k$-chains of simplicial complex $K$ together with the addition operation on $\mathbb{Z}$ constructs a free Abelian group $C_k(K)$, called chain group. To link chain groups from different dimensions, the $k$-boundary operator, $\partial_k: \ C_k(K)\rightarrow C_{k-1}(K)$, maps a $k$-chain in the form of a linear combination of $k$-simplexes to the same linear combination of the boundaries of the $k$-simplexes. For a simple example where the $k$-chain has one oriented $k$-simplex spanned by $k+1$ vertices as defined in Eq. (\ref{eq1}), its boundary operator is defined as the formal sum of its all $(k-1)$-faces:
\begin{equation}
\partial_k\sigma^k=\sum_{i=0}^k(-1)^i\sigma_{i}^{k-1}=\sum_{i=0}^k(-1)^i\left[v_0,\cdots, \hat{v_i} ,\cdots,\ v_k\right],
\end{equation}
where $\sigma_{i}^{k-1}=\left[v_0,\cdots, \hat{v_i} ,\cdots,\ v_k\right]$ is the $(k-1)$-simplex with its vertex $v_i$ being removed. The most important topological property is that a boundary has no boundary: $\partial_{k-1}\partial_{k}=\emptyset$.

A sequence of chain groups connected by boundary operators defines the chain complex:
\begin{equation}
\label{eq_chain}
\cdots	\xrightarrow{\partial_{n+1}}C_n(K)\xrightarrow{\partial_{n}}C_{n-1}(K)\xrightarrow{\partial_{n-1}}\cdots\xrightarrow{\partial_{1}}C_0(K)\xrightarrow{\partial_{0}}\emptyset.
\end{equation}
When $n$ exceeds the dimension of $K$, $C_n(K)$ is an empty vector space and the corresponding boundary operator is a zero map. 

\subsection{Filtration for multiscale chain complexes}

Filtration is a process that constructs a nested sequence of simplicial complex allowing a multiscale analysis of the point cloud. It creates a family of simplicial complexes ordered by inclusion (\autoref{fig1}c):
\begin{equation}
\emptyset= K^{t_0}\subseteq K^{t_1}\subseteq \cdots \subseteq K^{t_n}=K.
\label{eq2}
\end{equation}
where $K$ is the largest simplicial complex can be obtained from the point cloud. 

The filtration  induces a sequence of chain complexes
\begin{equation}
\begin{aligned}
&\cdots \xrightleftharpoons[\partial_{k+2}^{t_1*}]{\partial_{k+2}^{t_1}} &C_{k+1}^{t_1}& \xrightleftharpoons[\partial_{k+1}^{t_1*}]{\partial_{k+1}^{t_1}} &C_{k}^{t_1}& \xrightleftharpoons[\partial_{k}^{t_1*}]{\partial_{k}^{t_1}} \cdots \xrightleftharpoons[\partial_{1}^{t_1*}]{\partial_{1}^{t_1}} &C_{0}^{t_1}&\xrightleftharpoons[\partial_{0}^{t_1*}]{\partial_{0}^{t_1}}\emptyset\\
&&| \cap&&| \cap&&| \cap&\\
&\cdots \xrightleftharpoons[\partial_{k+2}^{t_2*}]{\partial_{k+2}^{t_2}} &C_{k+1}^{t_2}& \xrightleftharpoons[\partial_{k+1}^{t_2*}]{\partial_{k+1}^{t_2}} &C_{k}^{t_2}&\xrightleftharpoons[\partial_{k}^{t_2*}]{\partial_{k}^{t_2}} \cdots \xrightleftharpoons[\partial_{1}^{t_2*}]{\partial_{1}^{t_2}} &C_{0}^{t_2}&\xrightleftharpoons[\partial_{0}^{t_2*}]{\partial_{0}^{t_2}}\emptyset\\
&&| \cap&&| \cap&&| \cap&\\
&&\vdots&&\vdots&&\vdots&\\
&&| \cap&&| \cap&&| \cap&\\
&\cdots \xrightleftharpoons[\partial_{k+2}^{t_n*}]{\partial_{k+2}^{t_n}} &C_{k+1}^{t_n}&\xrightleftharpoons[\partial_{k+1}^{t_n*}]{\partial_{k+1}^{t_n}} &C_{k}^{t_n}&\xrightleftharpoons[\partial_{k}^{t_n*}]{\partial_{k}^{t_n}} \cdots \xrightleftharpoons[\partial_{1}^{t_n*}]{\partial_{1}^{t_n}} &C_{0}^{t_n}&\xrightleftharpoons[\partial_{0}^{t_n*}]{\partial_{0}^{t_n}}\emptyset\\
\end{aligned}
\end{equation}
where $C_k^t=C_k(K^t)$ is the chain group for subcomplex $K^t$, and its $k$-boundary operator is $\partial_k^t: C_k(K^t)\rightarrow C_{k-1}(K^t)$. $\partial_k^t$ is the co-boundary operator. Associated with the $k$-boundary operator, its adjoint operator is the $k$-adjoint boundary operator, $\partial_k^{t*}: C_{k-1}(K^t)\rightarrow C_{k}(K^t)$.

There are various simplicial complex that can be used to construct the filtration, such as Rips complex, $\check{C}$ech complex, and Alpha complex. For example, the Rips complex of $K$ with radius $t$ consists of all simplexes with diameter at most $2t$:
\begin{equation}
V(t)=\left\{\sigma\subseteq K\vert \mbox{diam} \ (\sigma)\leq 2t\right\}. 
\end{equation}

\subsection{Homology group and persistent homology}
With the chain complex defined in Eq. (\ref{eq_chain}), the $k$-cycle and $k$-boundary groups are defined as:
\begin{equation}
\begin{aligned}
Z_k={\rm ker} ~\partial_k=\{c\in C_k \mid \partial_k c=0\}\\
B_k={\rm im} ~\partial_{k+1}= \{ \partial_{k+1} c \mid c\in C_{k+1}\}\\
\end{aligned}
\end{equation}
Then the $k$-th homology group $H_k$ is defined as
\begin{equation}
H_k = Z_k / B_k.
\end{equation}
The $k$-th Betti number, $\beta_k$, is defined by the rank of $k$-th homology group $H_k$ which counts $k$-dimensional holes. For example, $\beta_0\!=\!{\rm rank}(H_0)$ reflects the number of connected components,
$\beta_1\!=\!{\rm rank}(H_1)$ reflects the number of loops,
and $\beta_2\!=\!{\rm rank}(H_2)$ reveals the number of voids or cavities.

Persistent homology is devised to track the multiscale topological information along the filtration \cite{edelsbrunner2000topological}. The inclusion map $K_i \subseteq K_j$ induces a homomorphism $f^{i,j}_k$ between homology groups $H_k(K_{t_i}) \to H_k(K_{t_j})$ for each dimension $k$. The $p$-persistent $k$-th homology {group of $K_t$} is defined by
\begin{equation}
H_k^{t,p} = Z^t_k/(B_k^{t+p}\cap Z^t_k),
	\label{eq_PH}
\end{equation}
where { $Z_k^t= {\rm ker} ~\partial_k^t$ and $B_k^{t+p} = {\rm im} ~\partial_{k+1}^{t+p}$}. Intuitively, this homology group records the $k$-dimensional homology classes of $K_{t}$ that are persistent at least until $K_{t+p}$. The birth and death of homology classes can be represented by a barcode, a set of intervals (\autoref{fig1}d).

\subsection{Combinatorial Laplacian.} 
For $k$-boundary operator $\partial_k: C_k\rightarrow C_{k-1}$ in $K$, let $\mathcal{B}_k$ be the matrix representation of this operator relative to the standard bases for $C_k$ and $C_{k-1}$ in $K$. $\mathcal{B}_k\in\mathbb{Z}^{M\times N}$ is the matrix representation of boundary operator under the standard bases $\left\{\sigma^k_i\right\}_{i=1}^N$ and $\left\{\sigma^{k-1}_j\right\}_{j=1}^M$ of $C_k$ and $C_{k-1}$.  Associated with the boundary operator $\partial_k$, the adjoint boundary operator is $\partial^*_{k}:C_{k-1}\rightarrow C_{k}$, where its matrix representation is the transpose of the matrix, $\mathcal{B}^T$, with respect to the same ordered bases to the boundary operator.

The $k$-combinatorial Laplacian, a topological Laplacian, is a linear operator $\Delta_k: C_k(K)\rightarrow C_k(K)$
\begin{equation}
\Delta_k:=\partial_{k+1}\partial_{k+1}^*+\partial^*_k\partial_k,
\end{equation}
and its matrix representation, $L_k$, is given by
\begin{equation}
L_k=\mathcal{B}_{k+1}\mathcal{B}_{k+1}^T+\mathcal{B}^T_k\mathcal{B}_k.
\end{equation}
In particular, the $0$-combinatorial Laplacian (i.e. graph Laplacian) is given as follows since $\partial_0$ is an zero map:
\begin{equation}
L_0=\mathcal{B}_1\mathcal{B}_1^T.
\end{equation}
The elements of $k$-combinatorial Laplaicn matrices are 
\begin{equation}
\left(L_k\right)_{i,j}=
\left\{
\begin{aligned}
&\mbox{deg}\left(\sigma^k_i\right),\mbox{ if }i=j\\
&1,\mbox{ if }i\neq j,\sigma^k_i\nsim_U\sigma^k_j\mbox{ and }\sigma^k_i\sim_L\sigma^k_j \mbox{ with similar orientation}\\
&-1, \mbox{ if }i\neq j,\sigma^k_i\nsim_U\sigma^k_j\mbox{ and }\sigma^k_i\sim_L\sigma^k_j \mbox{ with dissimilar orientation}\\
&0,\mbox{ if }i\neq j,\mbox{ either }\sigma^k_i\sim_U\sigma^k_j\mbox{ or }\sigma^k_i\nsim_L\sigma^k_j.\\
\end{aligned}
\right.
\end{equation}
For $k=0$, the graph Laplacian matrix $L_0$ is
\begin{equation}
\left(L_0\right)_{i,j}=
\left\{
\begin{aligned}
&\mbox{deg}\left(\sigma^0_i\right),\mbox{ if }i=j\\
&-1, \mbox{ if }i\neq j,\sigma^0_i\sim_U\sigma^0_j\\
&0,\mbox{ otherwise. }\\
\end{aligned}
\right.
\end{equation}

The multiplicity of zero spectra of $L_k$ gives the Betti-$k$ number, according to combinatorial Hodge theorem \cite{eckmann1944harmonische}:
\begin{equation}
\beta_k=\dim(L_k)-\mbox{rank}(L_k)=\mbox{null}(L_k).
\end{equation}
The Betti numbers describe topological invariants. Specifically, $\beta_0$, $\beta_1$, and $\beta_2$ may be regarded as the numbers of independent components, rings, and cavities, respectively. 

\subsection{Persistent spectral graph (PSG)}

The homotopic shape changes with a small increment of filtration parameter may be subject to noise from the data. The persistence may be considered to enhance the robustness when calculating the Laplacian. First, we define the $p$-persistent chain group $\mathbb{C}_k^{t,p}\subseteq C_k^{t+p}$ whose boundary is in $C^{t}_{k-1}$:
\begin{equation}
\mathbb{C}_k^{t,p}=\left\{\alpha\in C_k^{t+p}\ \vert \ \partial^{t+p}_k(\alpha)\in C_{k-1}^{t}\right\},
\end{equation}
where $\partial_k^{t+p}: \ C_k^{t+p}\rightarrow C_{k-1}^{t+p}$ is the $k$-boundary operator for chain group $C_k^{t+p}$. Then we can define a $p$-persistent boundary operator, $\eth_k^{t,p}$, as the restriction of $\partial_k^{t+p}$ on the $p$-persistent chain group $\mathbb{C}_k^{t,p}$:
\begin{equation}
\eth_k^{t,p}=\partial_k^{t+p}\vert_{\mathbb{C}_k^{t,p}}: \mathbb{C}_k^{t,p}\rightarrow C_{k-1}^t.S\end{equation}

Then PSG defines a family of $p$-persistent $k$-combinatorial Laplacian operators $\Delta_{k}^{t,p}: \ C_k(K_t)\rightarrow C_k(K_t)$ \cite{wang2020persistent, memoli2022persistent} which is defined as
\begin{equation}
\Delta_{k}^{t,p}=\eth_{k+1}^{t,p}\left(\eth_{k+1}^{t,p}\right)^*+\left(\partial_k^t\right)^*\partial_k^t.
\end{equation}
We denote $\mathcal{B}_{k+1}^{t,p}$ and $\mathcal{B}_k^t$ as the matrix representations for boundary operators $\eth_{k+1}^{t,p}$ and $\partial_{k}^{t}$, respectively. Then the Laplacian matrix for $\Delta_{k}^{t,p}$ is 
\begin{equation}
\mathcal{L}_k^{t,p}=\mathcal{B}_{k+1}^{t,p}\left(\mathcal{B}_{k+1}^{t,p}\right)^T+\left(\mathcal{B}_k^t\right)^T\mathcal{B}_k^t.
\end{equation}
Since the Laplacian matrix, $\mathcal{L}_k^{t,p}$, is positive-semidefinite, its spectra are all real and non-negative
\begin{equation}
S_k^{t,p}=\mbox{Spectra}(\mathcal{L}_k^{t,p})=\{(\lambda_1)_k^{t,p},(\lambda_2)_k^{t,p},\cdots,(\lambda_N)_k^{t,p}\},
\end{equation}
where $N$ is the dimension of a standard basis for $C_k^t$, and $\mathcal{L}_k^{t,p}$ has dimension $N\times N$. The $k$-persistent Betti number $\beta_k^{t,p}$ can be obtained from the multiplicity of harmonic spectra of $\mathcal{L}_k^{t,p}$:
\begin{equation}
\beta_k^{t,p}=\dim(\mathcal{L}_k^{t,p})-\mbox{rank}(\mathcal{L}_k^{t,p})=\mbox{null}(\mathcal{L}_k^{t,p})=\#\{i\vert (\lambda_i)_k^{t,p} \in S_k^{t,p}, \mbox{ and } (\lambda_i)_k^{t,p}=0\}.
\end{equation}
In addition, the rest of the spectra, i.e., the non-harmonic part, capture additional  geometric information. The family of spectra of the persistent Laplacians reveals the homotopic shape evolution \cite{qiu2023persistent}.

\newpage
\bibliographystyle{unsrtnat}